\documentclass[review]{elsarticle}

\usepackage{lineno,hyperref,makecell,amssymb,amsmath}
\usepackage{amstext,amsfonts}
\usepackage{graphicx,subfigure}
\usepackage{caption}
\usepackage{subfigure}
\usepackage{color}
\usepackage{pdfpages}
\usepackage{geometry}
\usepackage{float}
\modulolinenumbers[5]

\journal{arXiv:2110.01796}

\begin{document}
\captionsetup[figure]{labelfont={bf},name={Fig.},labelsep=period}
\begin{frontmatter}

\title{Gain stabilization and consistency correction approach for multiple SiPM-based gamma-ray detectors on GECAM}

\author[a]{Dali Zhang\corref{mycorrespondingauthor}}
\cortext[mycorrespondingauthor]{Corresponding author}
\ead{zhangdl@ihep.ac.cn}

\author[a]{Xinqiao Li}
\author[a]{Xiangyang Wen}
\author[a]{Shaolin Xiong}
\author[a]{Zhenghua An}
\author[a]{Yanbing Xu}
\author[b]{Xilei Sun}
\author[a]{Rui Qiao}
\author[a]{Zhengwei Li}
\author[a]{Ke Gong}
\author[a]{Dongya Guo}
\author[a]{Dongjie Hou}
\author[a]{Yanguo Li}
\author[a]{Xiaohua Liang}
\author[a]{Xiaojing Liu}
\author[a]{Yaqing Liu}
\author[a]{Wenxi Peng}
\author[a]{Sheng Yang}
\author[a]{Fan Zhang}
\author[a]{Xiaoyun Zhao}
\author[a]{Chao Zheng}
\author[a]{Chaoyang Li}
\author[a]{Qibin Yi}
\author[a]{Jiacong Liu}
\author[a]{Shuo Xiao}
\author[a]{Ce Cai}
\author[a]{Chenwei Wang}

\address[a]{Key Laboratory of Particle Astrophysics, Institute of High Energy Physics, Chinese Academy of Sciences, Beijing, China}
\address[b]{State Key Laboratory of Particle Detection and Electronics, Institute of High Energy Physics, Chinese Academy of Sciences, Beijing, China}

\begin{abstract}
Each satellite of the Gravitational wave high-energy Electromagnetic Counterpart All-sky Monitor (GECAM, mission) consists of 25 SiPM-based gamma-ray detectors (GRDs). {\color{black}SiPM based GRD benefit from being compact in size and low bias-voltage. However, the drift of the GRD gain with temperature is a severe problem for GRD performance.} An adaptive voltage supply source was designed to automatically adjust the SiPM bias voltage to compensate the temperature effects and keep the gain stable. {\color{black}This approach proved to be effective during both the on-ground and in-flight tests. }The in-flight measured variation of the GRD gain is within 2\%. To reduce the {\color{black}non-uniformity gain} of GRDs, an iterative bias voltage adjustment approach is proposed and implemented. The gain non-uniformity is reduced from 17\% to 0.6\%. In this paper, the gain stabilization and consistency correction approach are presented and discussed in detail.
\end{abstract}


\end{frontmatter}

\linenumbers
\section{Introduction}
\par
\begin{figure}[htbp]
  \centering
  \includegraphics[width=6 cm]{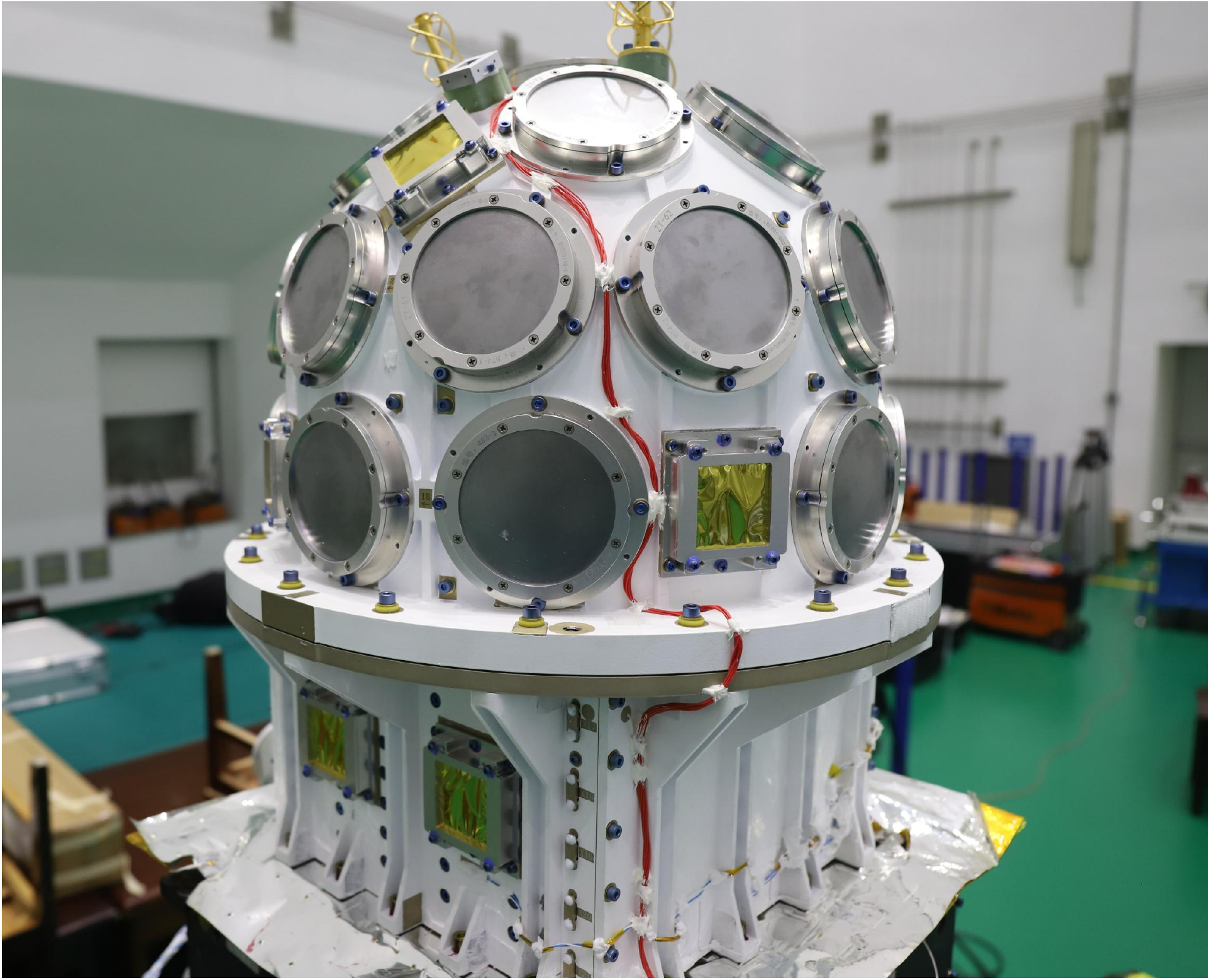}
  \hspace{1cm}
  \includegraphics[width=6.3 cm]{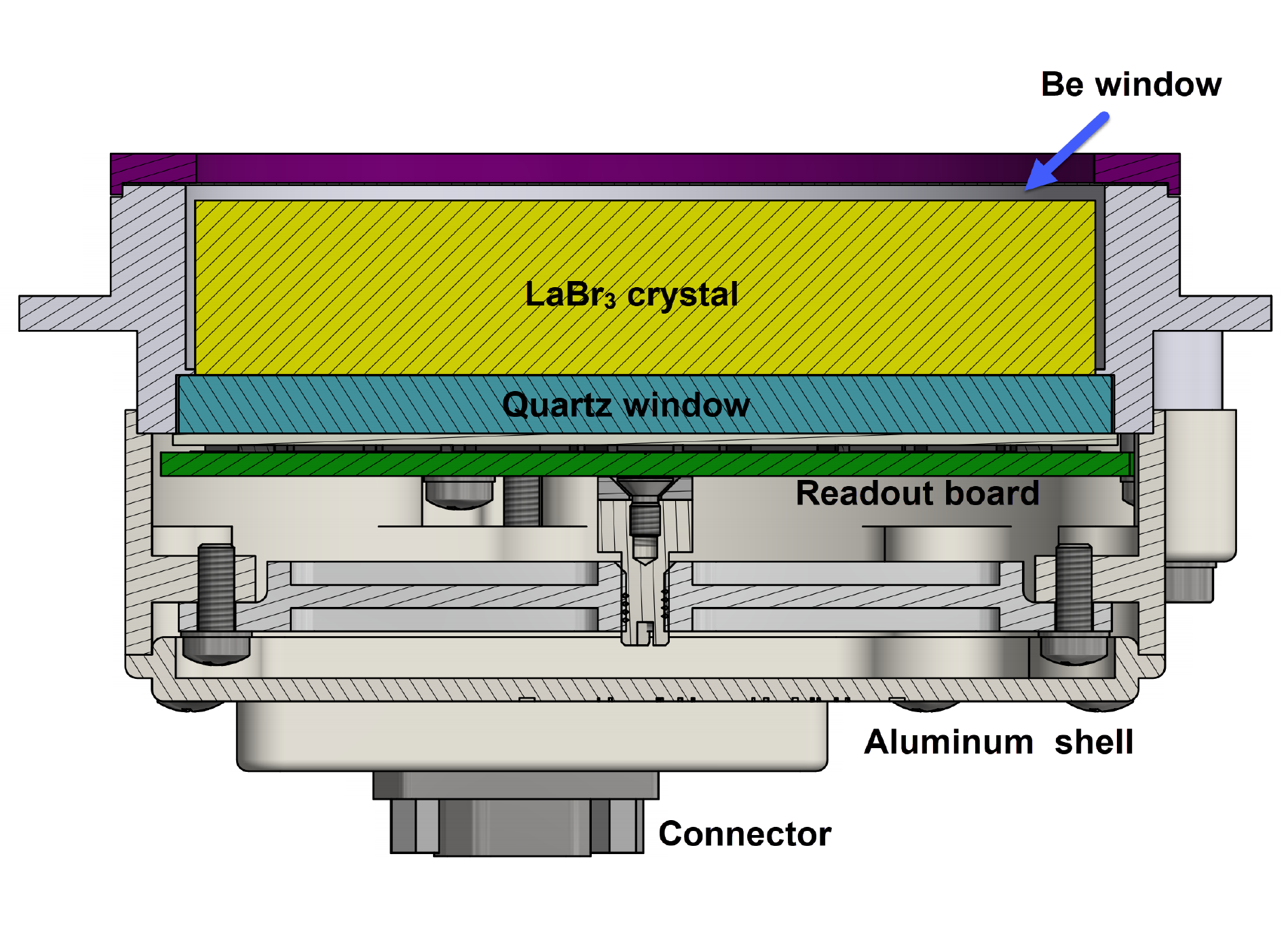}
  \caption{Left panel: Picture of the GECAM payload. Right panel: GRD structure.}\label{Fig1}
\end{figure}

\begin{figure}[htbp]
\vspace{-0.8cm}
  \centering
  \includegraphics[width=6 cm]{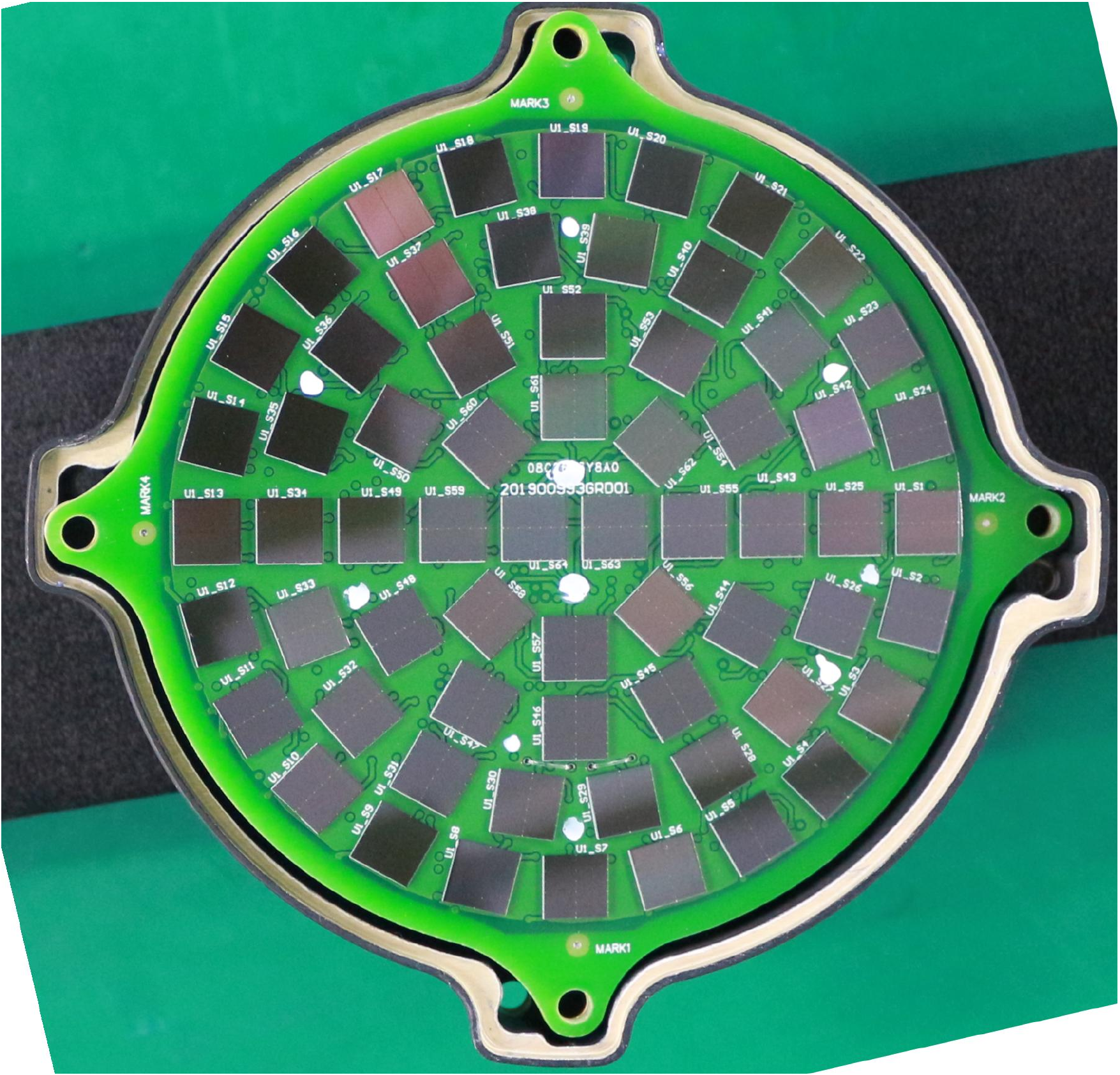}
  \hspace{1cm}
  \includegraphics[width=6 cm]{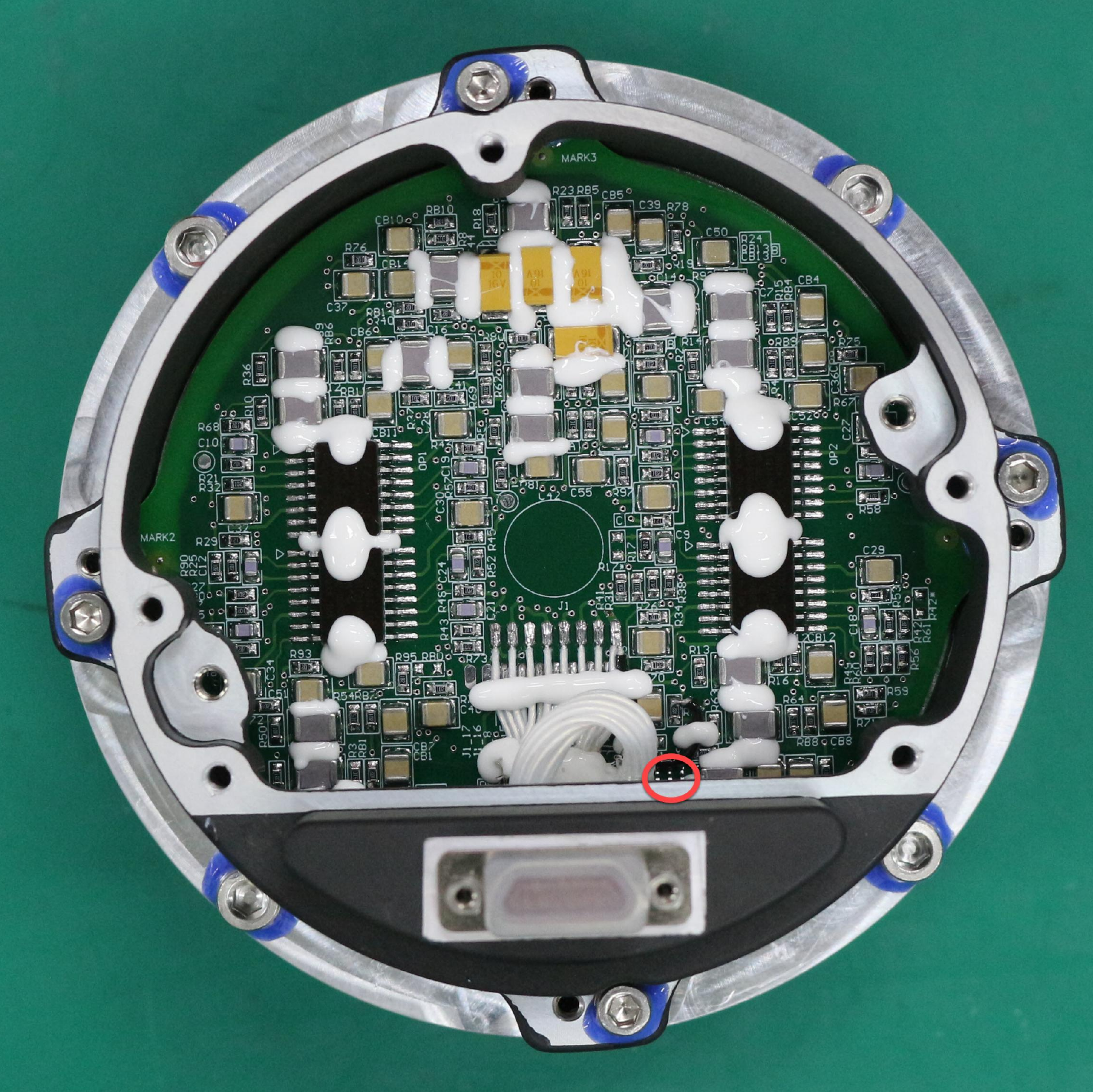}
  \caption{Left panel: Front view of GRD readout board. Right panel: Back view of GRD readout board. {\color{black}The temperature monitor chip is marked with red circle}}\label{Fig2}
\end{figure}

 {\color{black}On August 17, 2017, LIGO and Virgo jointly observed a gravitational wave (GW170817) originating from a merger of two neutron stars \cite{LIGO-Virgo}. At 1.7 seconds after the merger event, the Gamma-ray Burst Monitor (GBM) onboard Fermi \cite{Fermi-GBM} independently observed a gamma-ray burst (GRB 170817A) incident with the gravitational wave event. There are big challenges for electromagnetic counterparts detection because of lack of future gamma-ray missions. The Gravitational Wave Electromagnetic Counterpart All-sky Monitor (GECAM) \cite{GECAM-OV} was proposed in 2016 and launched on December 10, 2020 (Beijing Time). GECAM was designed to monitor GRBs, especially those associated with gravitational waves and guide follow-up observations in other electromagnetic wavelength such as radiation in the ultraviolet, infrared, soft X-rays, and radio bands \cite{X-ray-emission1},\cite{X-ray-emission2},\cite{Multi-messenger}. The orbit height of GECAM is 600 kilometers with an inclination angle of 29$^{\circ}$. The designed sensitivity of GECAM is about 2${\times}$10$^{-8}$erg/cm$^{2}$/s and the temporal resolution is 0.1 µs \cite{ GECAM-time}. The error of joint localization of SGR J1935+2154 by GECAM and GBM can be less than 0.4 degrees \cite{ GECAM-Localization},\cite{ GECAM-GCN}. The in-flight gamma burst trigger of GECAM can be promptly downlinked to the ground through the short message service of BeiDou Navigation Satellite System with a latency within 1 minute \cite{GECAM_localization}. GECAM consists of two micro-satellites and GECAM-B works for about 11 hours a day, while GECAM-A is temporarily not working yet.} As shown in Fig. \ref{Fig1} (left panel), each GECAM satellite consists of 25 {\color{black}circular}-shaped gamma-ray detectors (GRD) and 8 square-shaped charged particle detectors (CPD). {\color{black}GRD is composed of LaBr$_{3}$ crystal and a readout electronic board \cite{my-GRD},\cite{Lv-GRD},\cite{SiPM_pre}(Fig. \ref{Fig1} (right panel))}. In Fig. \ref{Fig2}, a 64-{\color{black}channel} SiPM (SensL MICROFJ-60035-TSV-TR) array is installed {\color{black}at} the front of {\color{black}the} readout board. The SiPM array pre-amplifier and temperature monitor module are on the back of readout board. The pre-amplifier outputs are divided into high gain and low gain channels and the typical energy range are 5.9 keV$\backsim$350 keV and 90 keV$\backsim$4.3 MeV for high gain and low gain, respectively.
\par SiPM based detectors are gradually applied in space projects such as GECAM, Insight-HXMT \cite{Lzw-sipm-gain}, SIRI-1 \cite{SIRI}, GRID \cite{GRID} due to {\color{black}their} compact size, low operating voltage and high quantum efficiency. {\color{black}However, SiPMs have} an obvious drawback that its gain has {\color{black}a} temperature dependence. There are many researches on the SiPM temperature compensation in literature. The negative feedback loop adjustment concept was successfully applied on Insight-HXMT \cite{Lzw-sipm-gain}, but this design requires precise selection of zener diodes with a proper temperature dependence. Other types of temperature compensation designs are based on temperature monitoring and feedback electronic systems \cite{SiPM_temp1},\cite{SiPM_temp2}. These designs are used for single detector adjustment with fixed temperature compensation parameters and can hardly meet the in-flight adjustment requirements in long term operation. {\color{black}There will be many un-expected new working conditions after launch and during the operation of serval years. The working temperature, total SiPM current and SiPM gain may change and the overall performance will deteriorate due to the long period of in-orbit irradiation. Thus the update of temperature compensation parameters is necessary.}
\par Here we developed an adaptive power supply for GECAM to reduce the temperature dependence of SiPM. The bias voltage of each GRD is automatically adjusted according to {\color{black}flexible} temperature-bias-voltage look-up table (LUT). Dedicated gain temperature dependence tests of GRD {\color{black}was implemented for gain stabilization}. Another problem is, due to different doping types and manufacturing processes, the light yield of the LaBr$_{3}$ crystals of GRDs are quite different which would worsen the gain non-uniformity of GRDs. These non-uniformity problem can also be solved by the proposed gain consistency correction approach {\color{black}for each GRD}.

\par The GECAM in-flight triggering and localization system \cite{GECAM_localization} determines a gamma-ray burst trigger from the energy response of GRD. The gain stabilization and consistency of multiple GRDs are vital for in-flight triggering and localization. The feasibility of the gain correction has been demonstrated by the on-ground experiments and in-flight tests.

\section{Gain stabilization and consistency correction in on-ground tests }\label{section2}
\subsection{Adaptive voltage source of GRD}
\par
\begin{figure}
  \centering
  \includegraphics[width=7 cm]{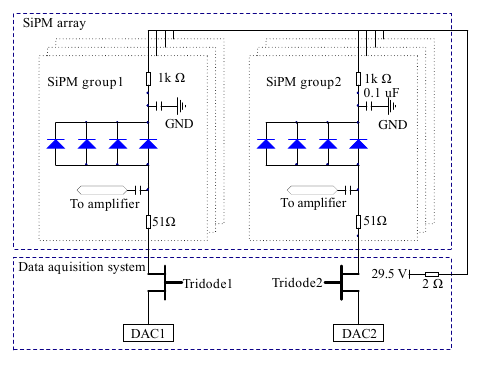}
  \hspace{0.1cm}
  \includegraphics[width=7 cm]{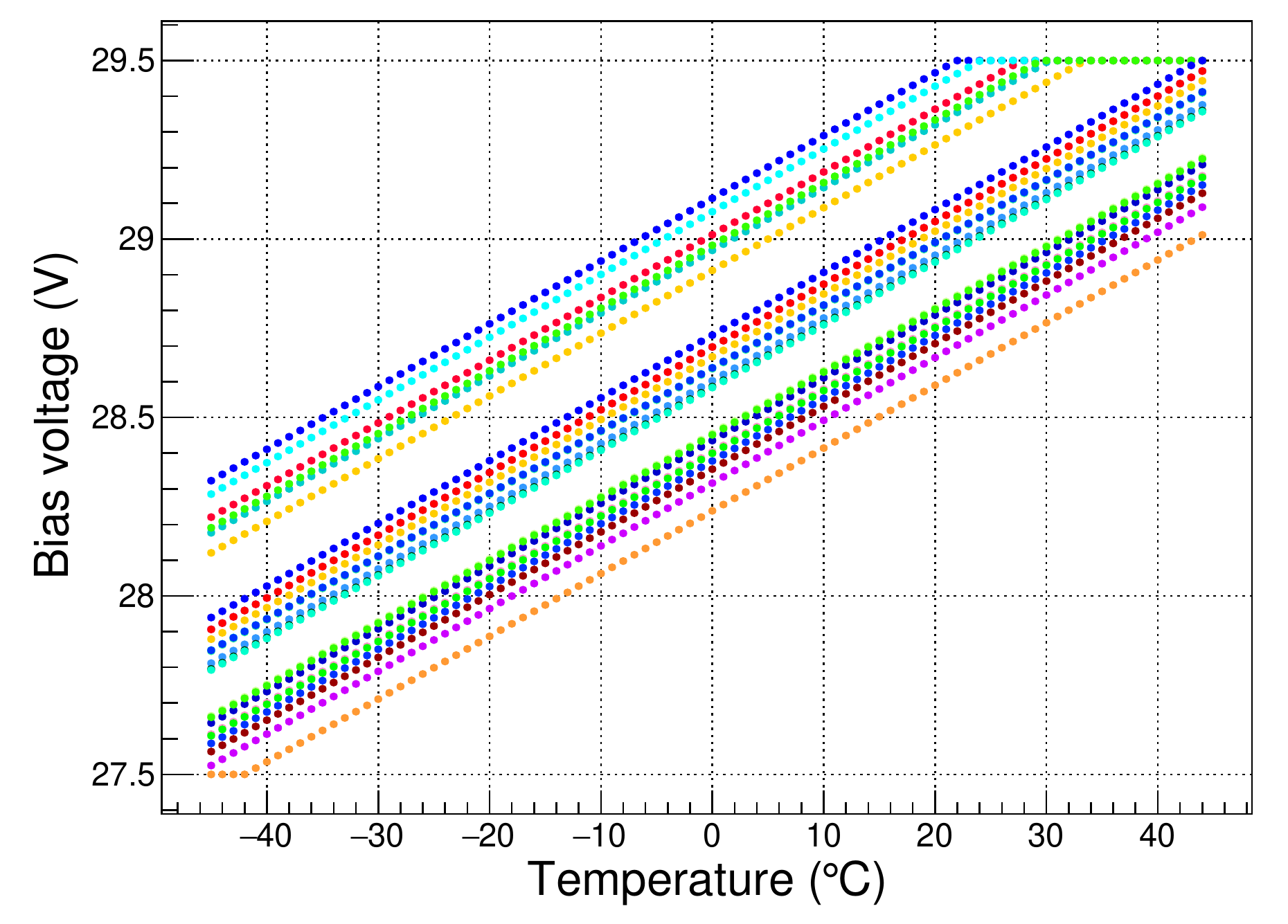}
  \caption{Left panel: Block scheme of SiPM array and power supply. Right panel: V10 version of bias voltage LUT.}\label{Fig3}
\end{figure}

 {\color{black}The main function of the adaptive voltage source is to stabilize the SiPM gain since it will significantly change with temperature variation. Because of the gain non-uniformity among the GRDs, the consistency correction is also achieved by adjusting the adaptive voltage source of each GRD.} Fig. \ref{Fig3} (left panel) shows the diagram of GRD SiPM array and the adaptive voltage source. All the SiPM cathodes are connected {\color{black}to} a 29.5 V voltage source. The anodes of the SiPM array are divided into two groups and each group contains 32 SiPM chips. Each group’s anodes are connected with the same emitter junctions of a triode that works as power switch. The collection junction of triode is connected {\color{black}to} a digital-to-analog convertor (DAC) chip. The DAC has a voltage output range of 0$\backsim$2V with an adjustment step of 0.5 mV. The SiPM bias voltage can be determined by subtracting {\color{black}the} voltage source (29.5 V) and the DAC output. Thus the adjustable bias voltage range is 27.5$\backsim$29.5V.
\par The payload data acquisition system reads the temperature monitor data once per second. If the temperature changes more than 0.5 $^{\circ}$C, the bias voltage will be refreshed. The bias voltage LUT covers a temperature range of -45$\backsim$45$^{\circ}$C for 25 GRDs. The bias voltage {\color{black}look-up table (LUT)} is a 25${\times}$80 array and it is stored in an refreshable EEPROM. Fig. \ref{Fig3} (right panel) shows the V10 version of bias voltage LUT where the points of various color represent different GRDs. {\color{black}Version number "V10" stands for the final bias voltage state before launch and the initial state in-flight.} Based on various in-flight operating conditions, different bias voltage LUT can be refreshed. The bias voltage adjustment principles are discussed in the following sections.

\subsection{Bias voltage and temperature dependence of GRD gain}\label{section2.2}
\par
\begin{figure}
\vspace{-0.2cm}
  \centering
  \includegraphics[width=7 cm]{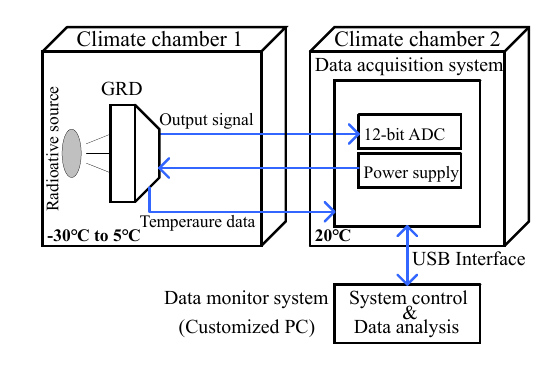}
  \hspace{0.1cm}
  \includegraphics[width=7 cm]{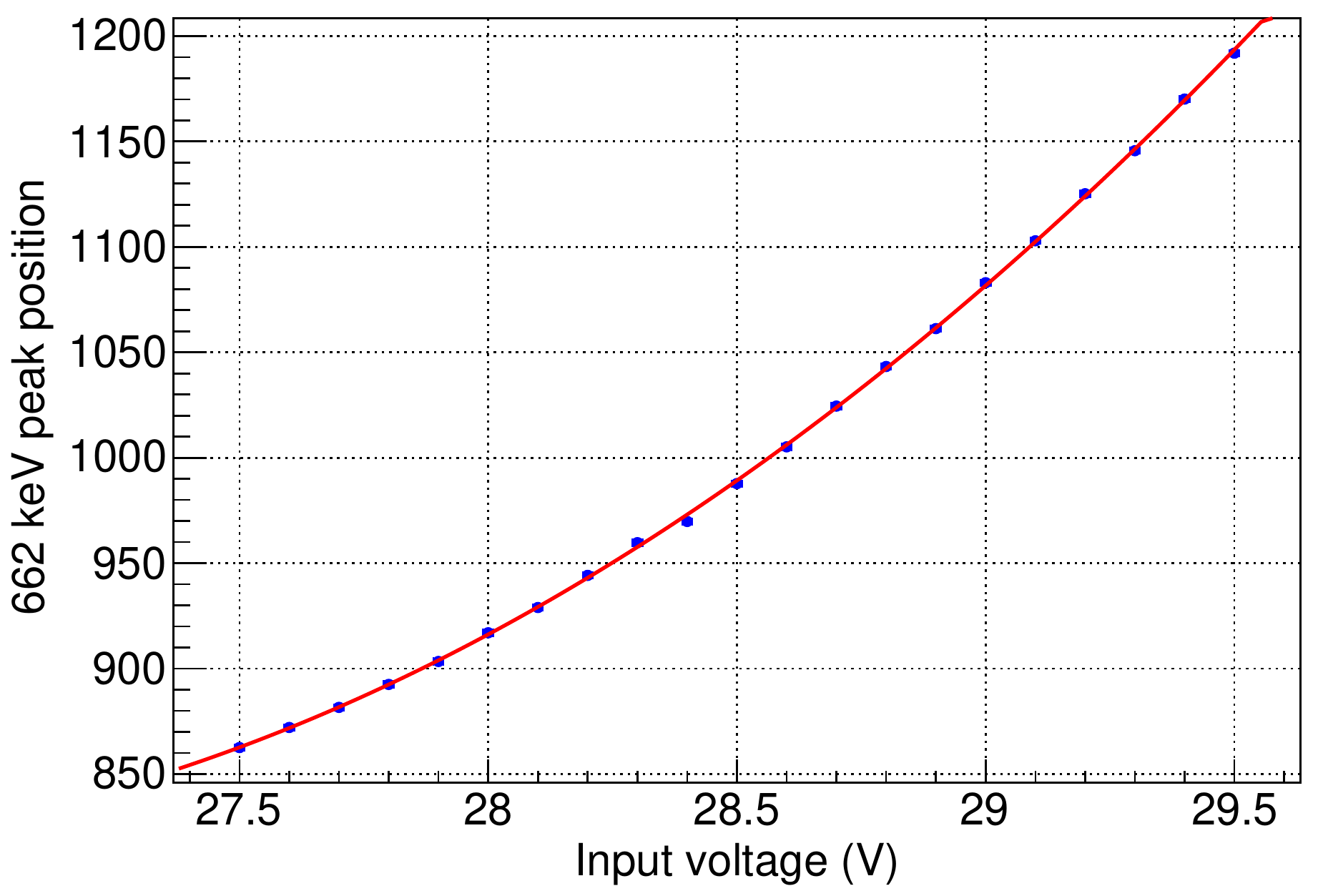}
  \caption{Left panel: Block diagram of the voltage and temperature dependence test setup. Right panel: Test results of GRD voltage dependence.}\label{Fig4}
\end{figure}

 The gain of a GRD is mainly affected by bias voltage and temperature variations of SiPM array. The voltage and temperature dependence tests are performed in a climate chamber. In Fig.\ref{Fig4} (left panel), a {\color{black}qualification test} GRD is placed in chamber 1 and the data acquisition system is in chamber 2. The temperature of GRD and data acquisition system is -20$^{\circ}$C and 20$^{\circ}$C, respectively. This temperature range includes the designed in-flight operating temperatures. The gain of a GRD is described by Eq.(\ref{GRD_gain}):
\begin{equation}
G_{det}=G_{sipm}\cdot PDE\cdot Ge\cdot LY\label{GRD_gain}
\end{equation}
\par where \textit{G$_{det}$}{\color{black}(ADC counts/keV)} is the total GRD gain. \textit{G$_{sipm}$} is the SiPM gain. \textit{PDE} is the SiPM photo detection efficiency. \textit{G$_{e}$} is the pre-amplifier gain and \textit{LY} is the light yield of LaBr$_{3}$ crystal. The SiPM gain can be described by Eq.(\ref{SiPM_gain}) \cite{SiPM_datasheet}, where \textit{V$_{ov}$} is the SiPM over voltage, \textit{C} is the microcell capacitance and \textit{q} is the electron charge. \textit{V$_{bias}$} and \textit{V$_{BR}$} are SiPM bias voltage and break down voltage, respectively.
\begin{equation}
G_{sipm}=\frac{C}{q}V_{ov}=\frac{C}{q}(V_{bias}-V_{BR})\label{SiPM_gain}
\end{equation}
\par The photo detection efficiency is proportional to overvoltage\cite{SiPM_datasheet} as shown in Eq.(\ref{PDE}):
\begin{equation}
PDE=a(V_{bias}-V_{BR})+b\label{PDE}
\end{equation}
\par Under stable temperature, the peak position of given gamma-ray energy can be described as Eq.(\ref{PP}). {\color{black} It shows that the peak position is jointly affected by the gain and PDE of SiPM. Since the gain and PDE are both linear to bias voltage, the voltage dependence can be fitted with a quadratic equation. The tested voltage dependence of GRD under -20$^{\circ}$C is drawn in Fig.\ref{Fig4} (right panel). The fit parameters are \textit{p$_{0}$}=27850, \textit{p$_{1}$}=-2050, \textit{p$_{1}$}=38.87 and $\chi$$^{2}$/ndf is 10.38/18.}
\begin{equation}
P=G_{det}\cdot E=G_{sipm}\cdot PDE\cdot Ge\cdot LY\cdot E=(p_{0}+p_{1}\cdot V_{bias} +p_{2}V_{bias}^{2})\cdot E\label{PP}
\end{equation}
\par
\begin{figure}[htbp]
  \centering
  \includegraphics[width=7.5 cm]{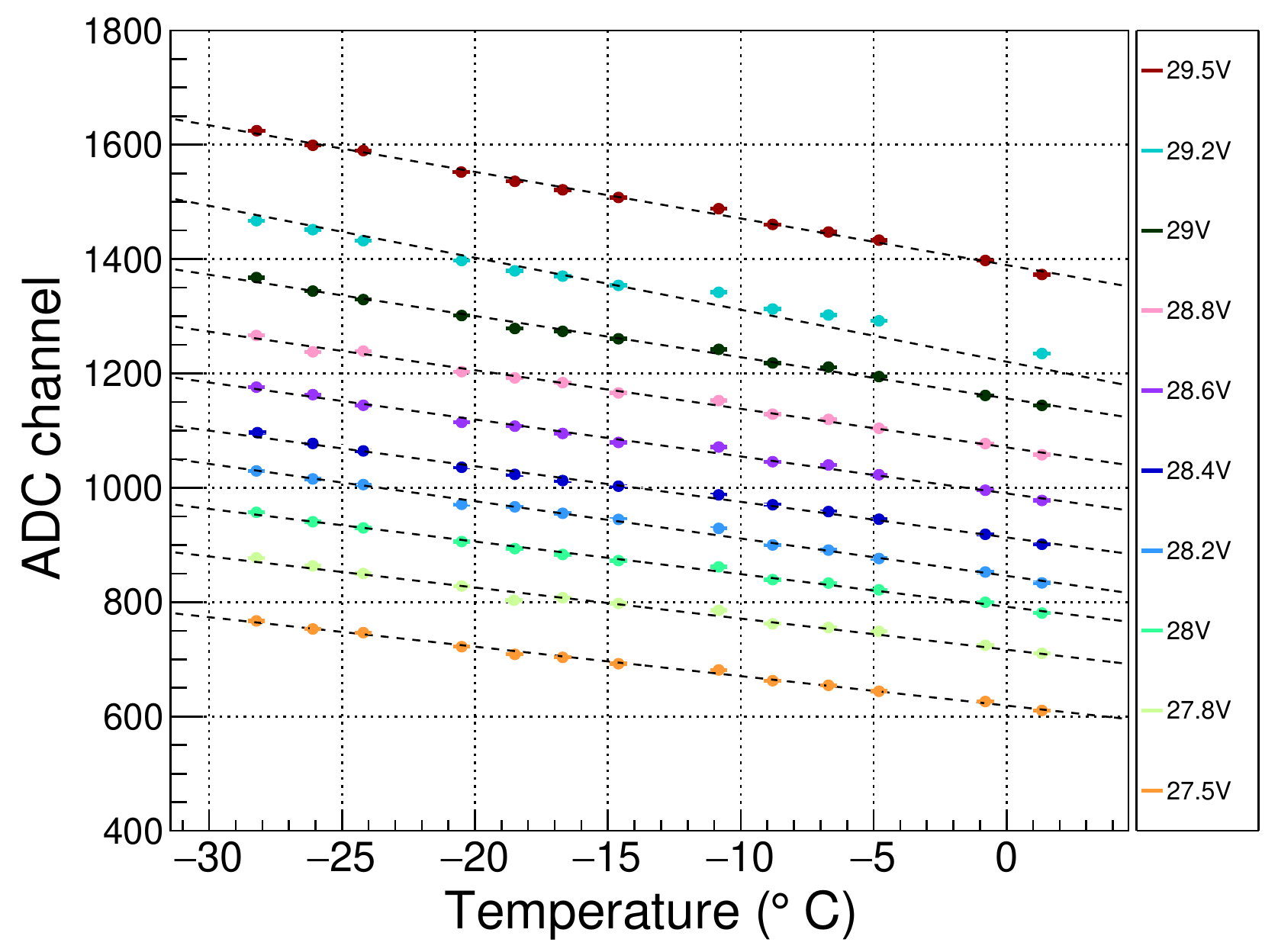}
  \caption{On-ground test results of GRD temperature dependence. The peak positions of 662 keV (Cs$^{137}$) under various bias voltage and temperature are measured.}\label{Fig5}
  \vspace{-0.8cm}
\end{figure}

 The tested temperature dependence is shown in Fig.\ref{Fig5}. The total temperature dependence mainly comes from the SiPM gain variation, while the temperature dependence of LaBr$_{3}$ crystal {\color{black}(0.01\%/$^{\circ}$C)} \cite{Crystal_temp} and the data acquisition system is {\color{black}relatively small}. The main cause of the SiPM gain drift with temperature is breakdown voltage(\textit{V$_{BR}$}) shift \cite{SiPM_datasheet}  shown in Eq.\ref{Vbr}:
\begin{equation}
\Delta V_{BR}=\alpha \cdot \Delta T \label{Vbr}
\end{equation}
\par As shown in Eq.(\ref{GRD_gain}), the total GRD gain is affected by SiPM gain and photo detection efficiency that change with temperature. Then the peak position will shift with temperature variation. {\color{black}The designed in-flight operating temperature for GRD is around -20 $^{\circ}$C and this test needs to cover this range. Although the climate chamber temperature is set to -30 to 0 $^{\circ}$C, the measured temperature is -28.2 to 1.3 $^{\circ}$C. In this range, measured GRD peak position variation is 20.1\% under a typical bias voltage of 28 V (Fig.\ref{Fig5}).} The temperature dependence $\alpha$ (17.75$\pm$0.25 mV/$^{\circ}$C) of bias voltage is extracted from the radioactive source measurements of spare GRDs. Because of the positive temperature dependence of LaBr$_{3}$ crystal \cite{Crystal_temp}, GRD has a smaller temperature dependence than single SiPM chip (21.5 mV/$^{\circ}$C) \cite{SiPM_datasheet}.
\subsection{Temperature compensation principle and on-ground gain stabilization}
\par
\begin{figure}[H]
  \centering
  \includegraphics[width=6 cm]{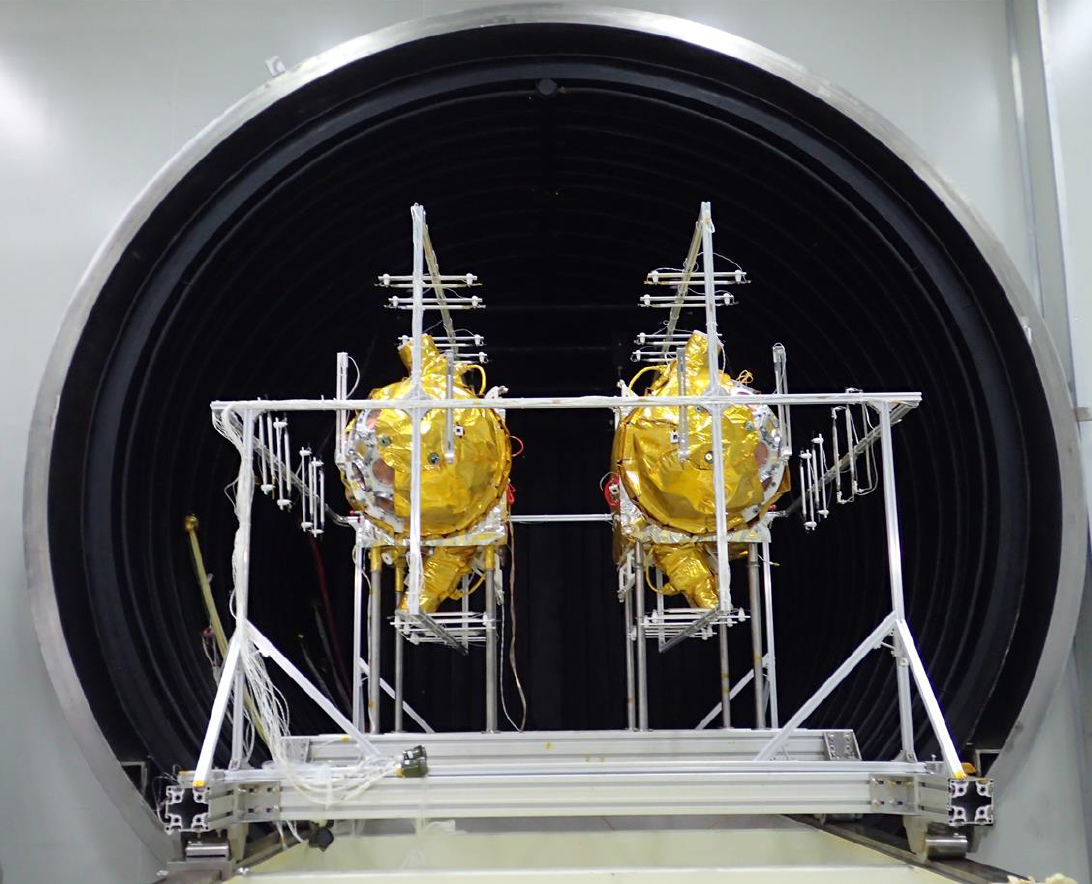}
  \hspace{0.5cm}
  \includegraphics[width=7 cm]{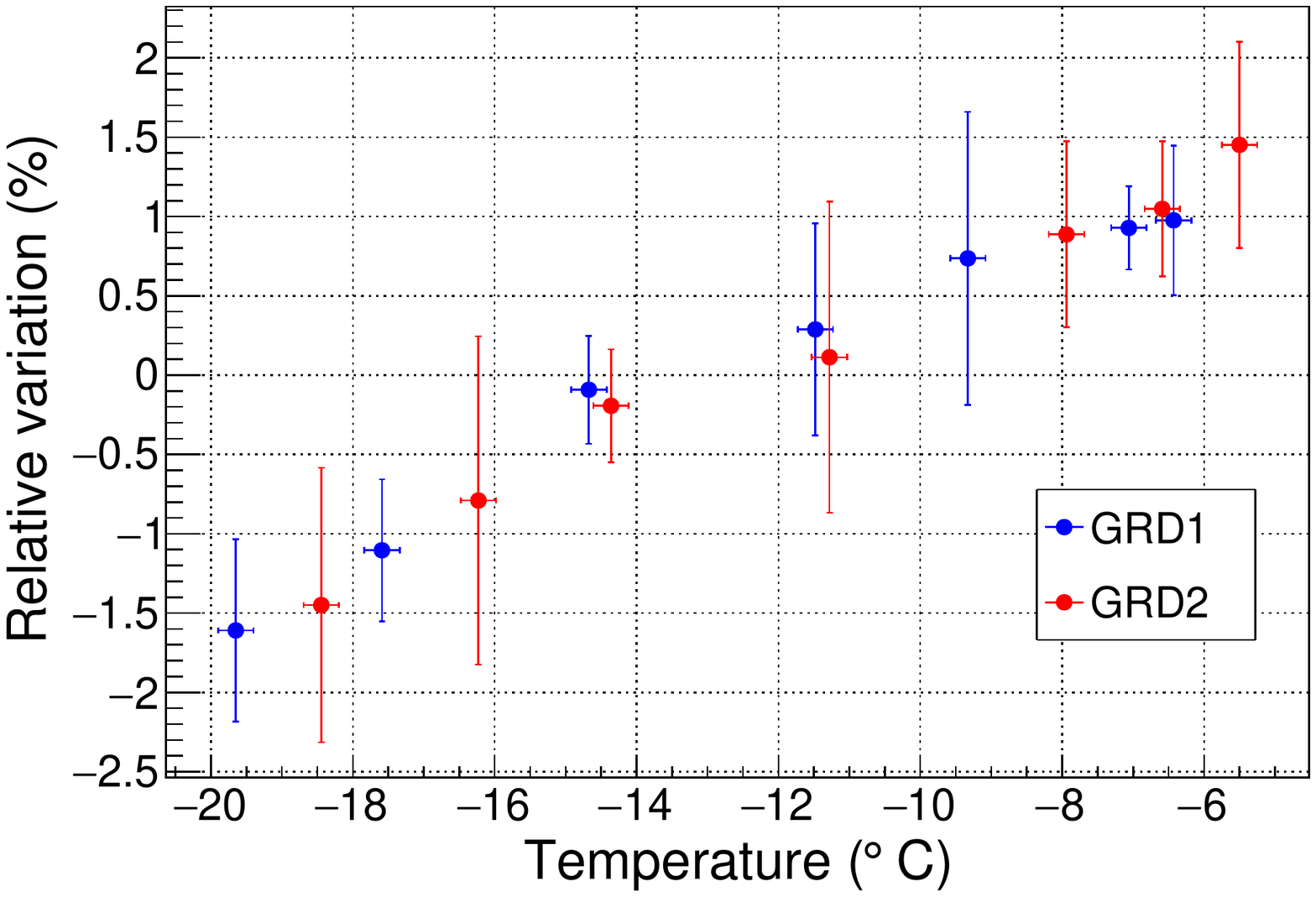}
  \caption{Left panel: Thermal vacuum experiment of GECAM. Two payloads are installed in a huge vacuum tank. Right panel: Relative variation from the mean value of the 37.4 keV peak position.}\label{Fig6}
\end{figure}

 During the integrated assembly tests of {\color{black}the} flight model in phase D, a thermal vacuum experiment of the whole payload (Fig.\ref{Fig6} left panel)  was implemented. To avoid potential damage of the {\color{black}flight} GRD in the thermal test, only a few {\color{black}flight qualification test} GRDs were installed on the satellite. The GRD data is acquired by the real satellite data transmission system. The approach to stabilize the gain is to keep a constant over voltage when temperature changes. The temperature compensation of GRD bias voltage is computed as Eq.(\ref{T_compen}):
\begin{equation}
V_{bias}=V_{R}+\alpha \cdot(T-T_{R}) \label{T_compen}
\end{equation}
\par where \textit{V$_{R}$} is 28 V, \textit{T$_{R}$} is 20 $^{\circ}$C and $\alpha$ is 18 mV/$^{\circ}$C, which comply with the operating parameters in on-ground calibration tests. \textit{T} is the temperature given by monitor chip {\color{black}on the marked area of readout board in Fig.\ref{Fig2} (right panel)}. Fig.\ref{Fig6} (right panel) shows the gain temperature dependence of the two installed GRDs. The relative LaBr$_{3}$ intrinsic 37.4 keV peak position variation with temperature remains stable to less than 1.6\% in the -20$^{\circ}$C$\backsim$-6$^{\circ}$C range. This variation shows the remaining residual from the mean value of all tested peak positions after temperature compensation and it is not perfectly stable. Further GRD gain stabilization results are presented in section \ref{section3.1}.
\subsection{Bias voltage adjustment principle of multiple GRDs and gain consistency}\label{section2.4}
\par
\begin{figure}[htbp]
  \centering
  \includegraphics[width=7.5 cm]{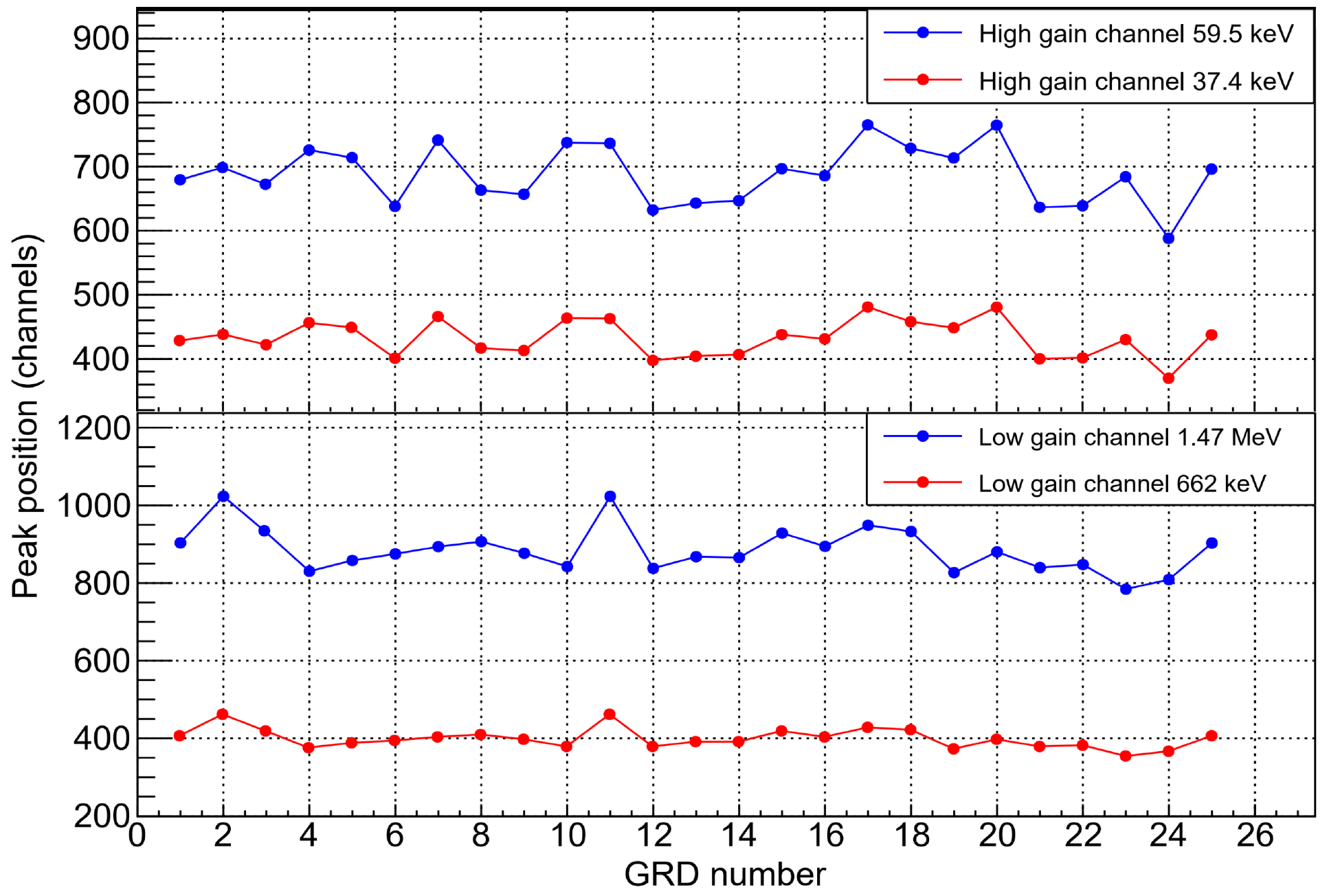}
  \caption{Gain non-uniformity of GRDs in on-ground calibration test. The peak positions of 37.4 keV and 1.47 MeV LaBr$_{3}$ intrinsic energy and radioactive source energy lines of 59.5 keV and 662 keV are plotted.}\label{Fig7}
\end{figure}

 The 25 GRDs on GECAM-B satellite can be categorized into two groups: Ce (5\%) doped and Ce (5\%)+Sr (1.5\%) co-doped LaBr$_{3}$. {\color{black}GRDs are made of LaBr$_{3}$ crystals of different raw materials and packaging process. Thus results in various LaBr$_{3}$ crystal light yield and gain no-uniformity of GRDs.} The on-ground radioactive source calibration test is performed under {\color{black}20$\pm$0.2 $^{\circ}$C} and the bias voltage of all tested GRDs is 28 V. As shown in Fig.\ref{Fig7}, the maximum non-uniformity of LaBr$_{3}$ intrinsic 1.47 MeV peak position is 17\%. The non-uniformity of GRDs are minimized by bias adjustment of SiPM arrays. For in-fight bias voltage adjustments, the {\color{black}updated} times of bias voltage LUT needs to be minimized and each update should be conservative so as to not affect the normal gamma-ray burst observation. Therefore, we designed a simplified and effective adjustment approach to meet these requirements above.
\par Based on the previous voltage dependence test in section \ref{section2.2}, to correct the GRD non-uniformity, the SiPM bias adjustment range is {\color{black}found to be 0.8 V}. In this small bias range, the SiPM quantum efficiency is approximately constant. Then the voltage-versus-peak position dependence can be expressed as Eq.(\ref{simple_L}), where the subscript i is the GRD number from 1 to 25:
\begin{equation}
P_{i}=a_{i}(V_{biasi}-V_{BRi})\cdot E+b_{i},a_{i}=PDE_{i}\cdot Ge_{i}\cdot LY_{i} \label{simple_L}
\end{equation}
\par
\begin{figure}[htbp]
  \centering
  \includegraphics[width=6 cm]{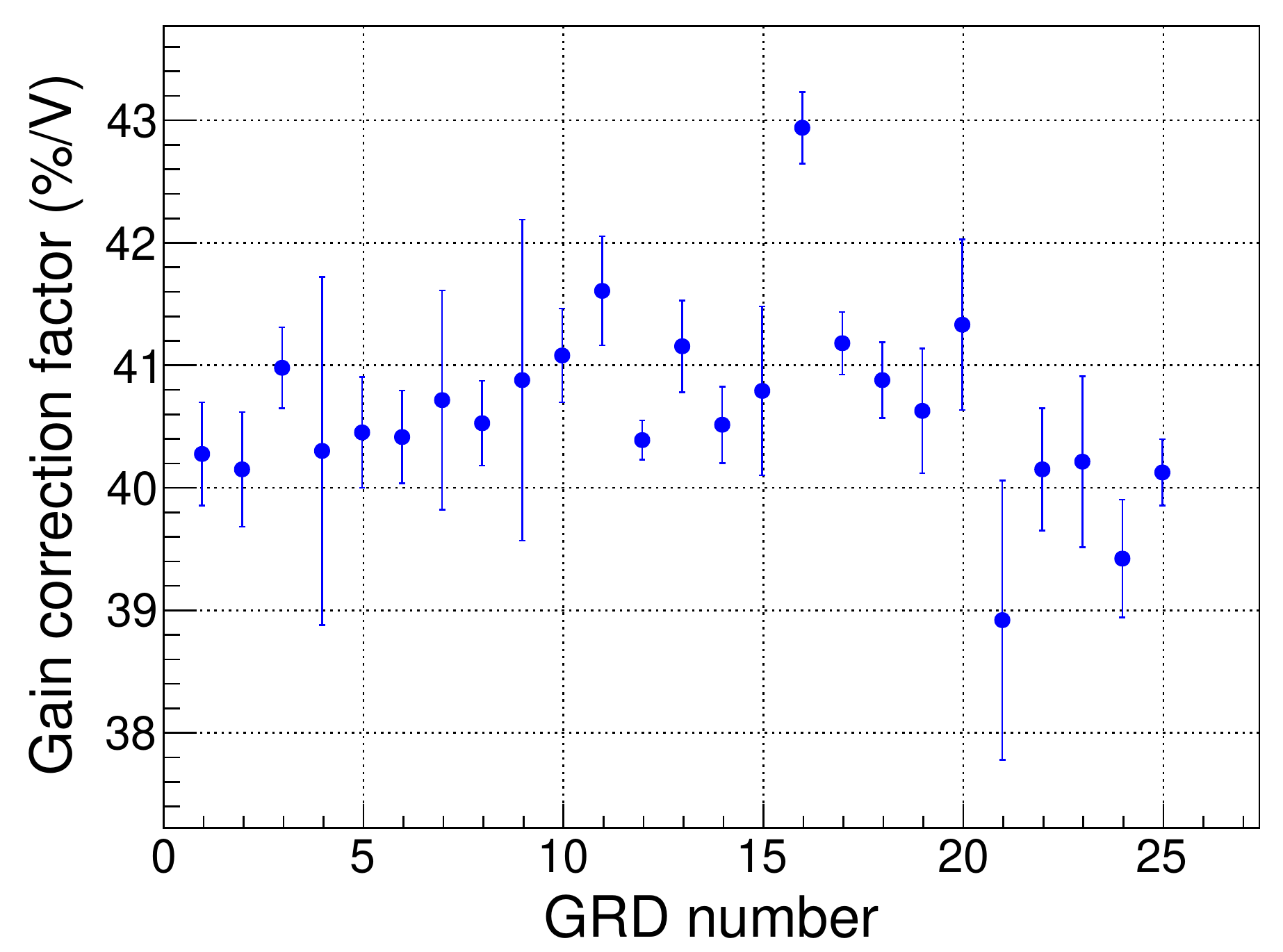}
  \caption{Measured gain correction factor \textit{F} of 25 GRDs on GECAM-B.}\label{Fig8}
\end{figure}

 Since the light yield of LaBr$_{3}$ crystal decreases with the increase of energy \cite{NPR}, the LaBr$_{3}$ intrinsic energy line of 1.47 MeV is used for the gain adjustment. Moreover, the selection of 1.47 MeV peak position alignment could result in a larger energy range where the peak position difference between different doped GRDs is very small {\color{black}(this will be discussed in section \ref{section3.2})}. In the multiple GRDs gain adjustment process, we define the reference operating parameters as \textit{P$_{R}$} (peak position of 1.47 MeV), \textit{V$_{R}$} (bias voltage of SiPM array) and \textit{T$_{R}$} (temperature of SiPM array). The \textit{P$_{R}$}, \textit{V$_{R}$} and \textit{T$_{R}$} are 1420 channels, 28 V and 20$^{\circ}$C, respectively. Based on the temperature dependence in Eq.(\ref{T_compen}) and linear voltage-versus-peak position dependence in Eq.(\ref{simple_L}), the gain correction factor \textit{F} of each GRD is defined as Eq.(\ref{G_corr}):
\begin{equation}
F(\%/V)= \frac{100 \cdot(P_{v1}-P_{v2})}{V_{ov1}-V_{ov2}}/P_{R}= \frac{100 \cdot(P_{v1}-P_{v2})}{(V_{bias1}-V_{bias2})-\alpha(T_{1}-T_{2})}/P_{R}\label{G_corr}
\end{equation}
\par where the subscript 1 and 2 stands for GRD parameters under \textit{V$_{bias1}$} and \textit{V$_{bias2}$}, respectively.  \textit{V$_{bias1}$} and \textit{V$_{bias2}$} are 27.5 V and 28.5 V, respectively. Temperature difference between GRDs is also corrected by using temperature dependence $\alpha$. Therefore, \textit{F} shows the relative gain-voltage dependence around the reference operating bias voltage and temperature. The measured \textit{F} of all 25 GRDs on the GECAM-B satellite are plotted in Fig.\ref{Fig8}.
\par Because substantial bias voltage adjustment in in-flight operation is risky, the gain adjustment of multiple GRDs is conservative and needs serval iterations to achieve desired results. The bias voltage of next iteration is defined as \textit{V$_{bias}$(i,n+1)} in Eq.(\ref{bias_table}) , where i is the GRD number and n is the number of iteration times.
\begin{equation}
V_{bias}(i,n+1)=V_{bias}(i,n)+(1-P(i,n)/P_{R})/F_{i}+\alpha(T(i,n)-T_{R})\label{bias_table}
\end{equation}
\par where \textit{V$_{bias}$(i, n)}, \textit{P(i, n)} and \textit{T(i, n)} represents the current bias voltage, measured peak position of 1.47 MeV and SiPM array temperature respectively. In the first iterative adjustment (\textit{n}=0), \textit{V$_{bias}$(i, 0)}=28 V. A temperature compensation is also performed in Eq.(\ref{bias_table}) by making use of the temperature dependence $\alpha$.
\par
\begin{figure}[htbp]
  \centering
  \includegraphics[width=7.6 cm]{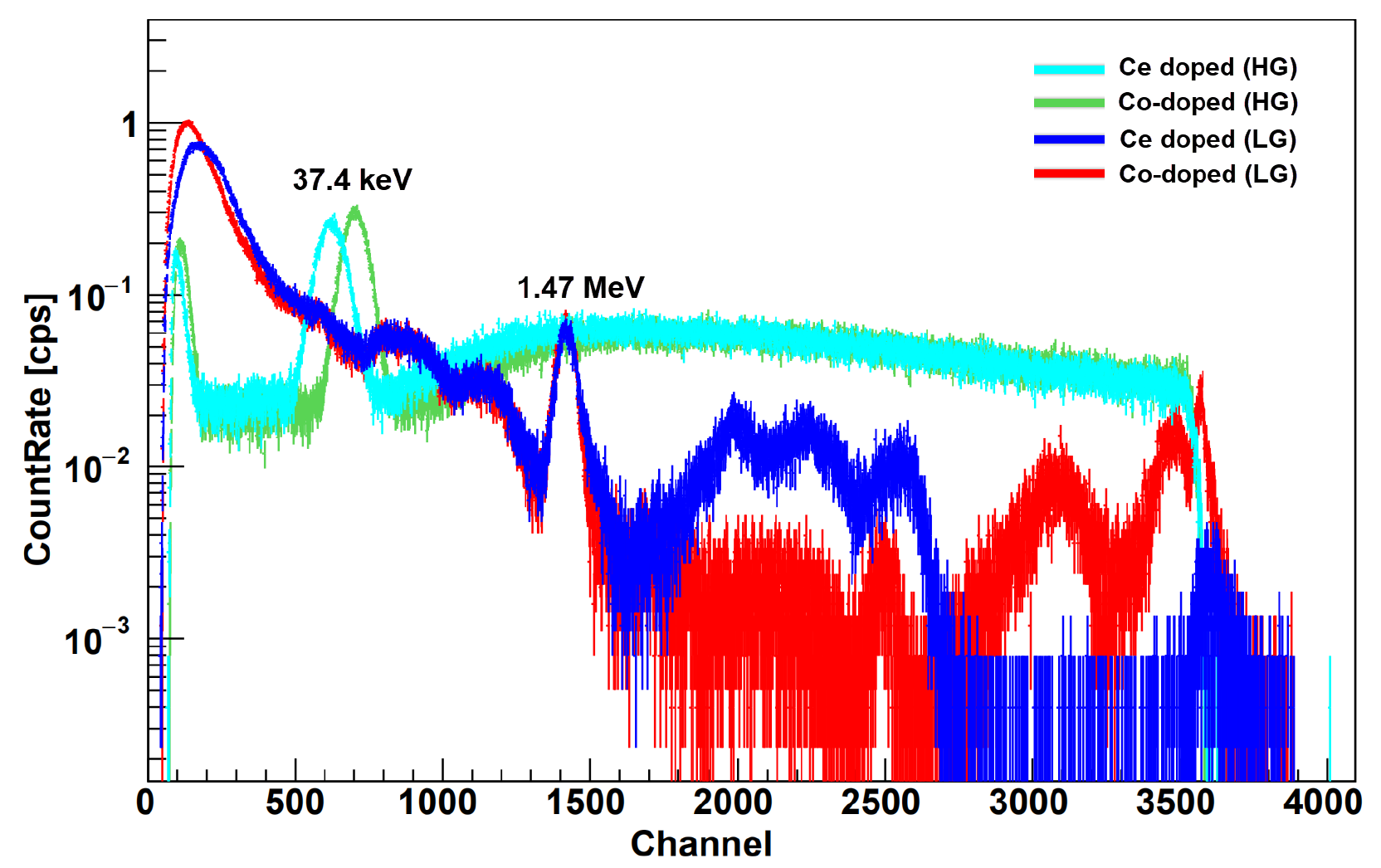}
  \hspace{0.1cm}
  \includegraphics[width=7 cm]{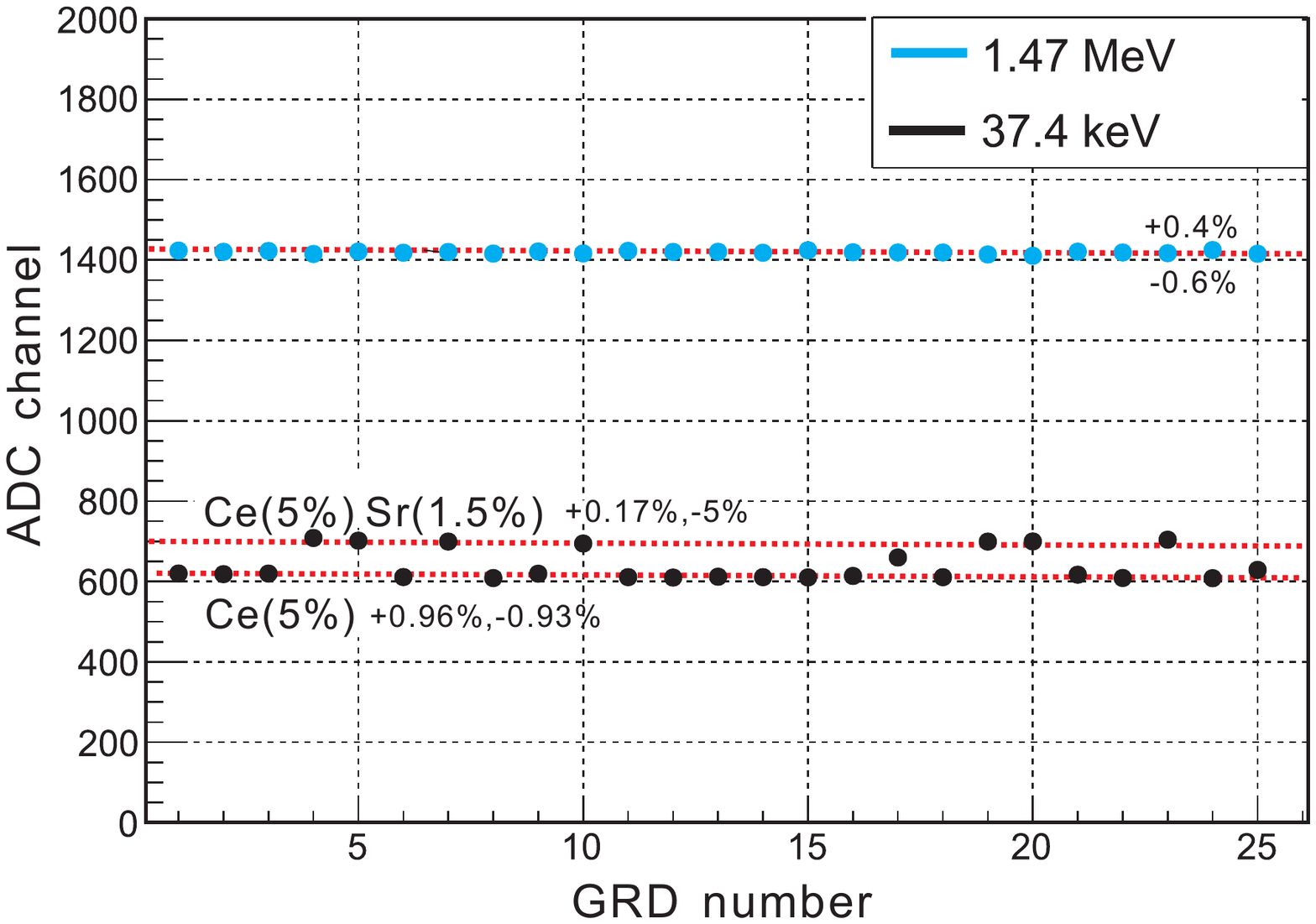}
  \caption{On-ground consistency correction results. Left panel: Measured intrinsic activity of different Ce doped GRD23 and co-doped GRD24. Right panel: Non-uniformity of GRDs under the V10 version of bias voltage LUT.}\label{Fig9}
\end{figure}

 After the bias voltage of each GRD (at temperature \textit{T$_{R}$)} is determined, the temperature-bias-voltage LUT: \textit{V$_{bias}$(i, T$_{i}$)=V$_{bias}$(i,T$_{R}$)+$\alpha$(T$_{i}$-T$_{R}$)} can be obtained. In the on-ground tests, the non-uniformity decreased from 17\% to less than 5\% in the first iteration, 2\% in the second iteration and 0.6\% in the third iteration. {\color{black}Each iteration is based on the previous measured data and results in a smaller deviation to the desired reference parameter (\textit{P$_{R}$}) in Eq.(\ref{bias_table}).} Fig.\ref{Fig9} (left panel) shows the background comparison of Ce doped and Ce+Sr co-doped GRDs after consistency correction. The non-uniformities are plotted in Fig.\ref{Fig9} (right panel). The 1.47 MeV peaks are highly aligned with 0.6\% max peak position variations from mean value, while the 37.4 keV intrinsic energy lines of co-doped LaBr$_{3}$ have a higher peak position.
\par
\begin{figure}[htbp]
  \centering
  \includegraphics[width=7.4 cm]{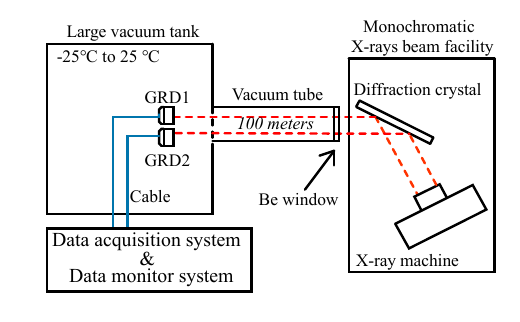}
  \hspace{0.1cm}
  \includegraphics[width=7 cm]{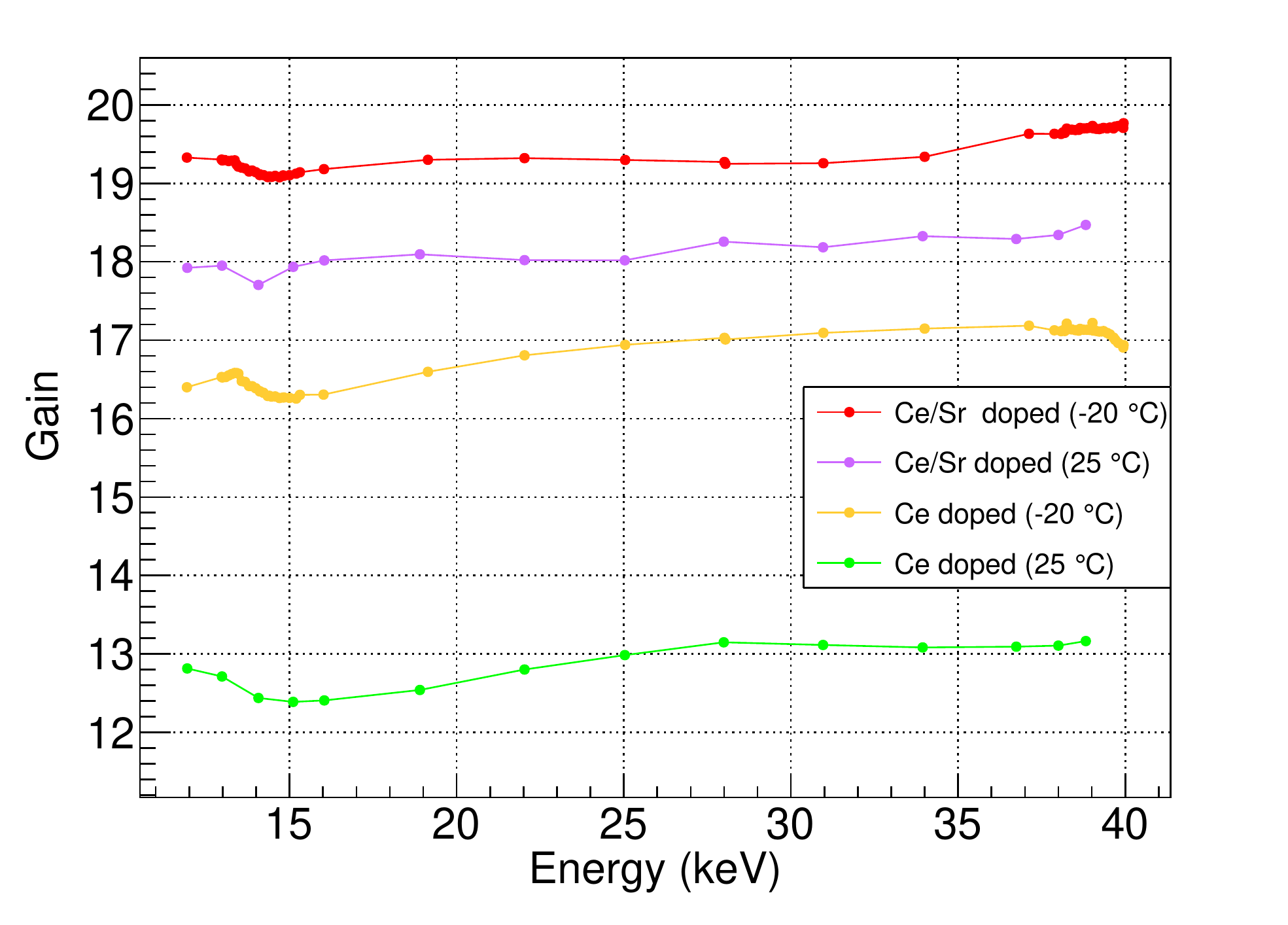}
  \caption{Left panel: Experimental setup of the GRD energy response in 100-m vacuum X-ray calibration facility. Right panel: Gain comparison between single-doped and co-doped GRD. The peak positions of 1.47 MeV are adjusted to 1420 channels for both of the GRDs under all temperature.}\label{Fig10}
  \vspace{-0.2cm}
\end{figure}

 To further understand the peak position difference of 37.4 keV intrinsic energy lines after consistency correction, we carried out a calibration experiment of two types of GRDs in the 100-m vacuum X-ray calibration facility at the Institute of High Energy Physics \cite{100m_facility} (Fig.\ref{Fig10} left panel). The (5$\backsim$40) keV monochromatic X-rays beam facility \cite{NIM_beam} was used to generate x-rays below 40 keV at the start of 100 m beam line. The Fig.\ref{Fig10} right panel shows the calibration experiment results of the GRD gain (channel/energy) difference \cite{LaBr_dope}. In this experiment, the same consistency correction is implemented. {\color{black}The measured peak positions of 1.47 MeV intrinsic energy line are aligned to 1427$\pm$10 channels under all temperatures.} Fig.\ref{Fig10} (right panel) shows that the gain of different doped LaBr$_{3}$ crystals changed with energy range and temperature \cite{LaBr_temp1},\cite{LaBr_temp2}. The co-doped GRD has a higher gain at 37.4 keV under the room temperature of 20 $^{\circ}$C and designed in-flight operation temperature of -20 $^{\circ}$C. This difference is also observed in in-flight results {\color{black}(shown in section \ref{section3.2}). Further} research on energy responds of this calibration test is described in a dedicated paper \cite{100m-beam-test}.
\section{In-flight gain stabilization and consistency}\label{section3}
\subsection{In-flight GRD gain stabilization with temperature variation}\label{section3.1}
\par
\begin{figure}[htbp]
  \centering
  \includegraphics[width=7.4 cm]{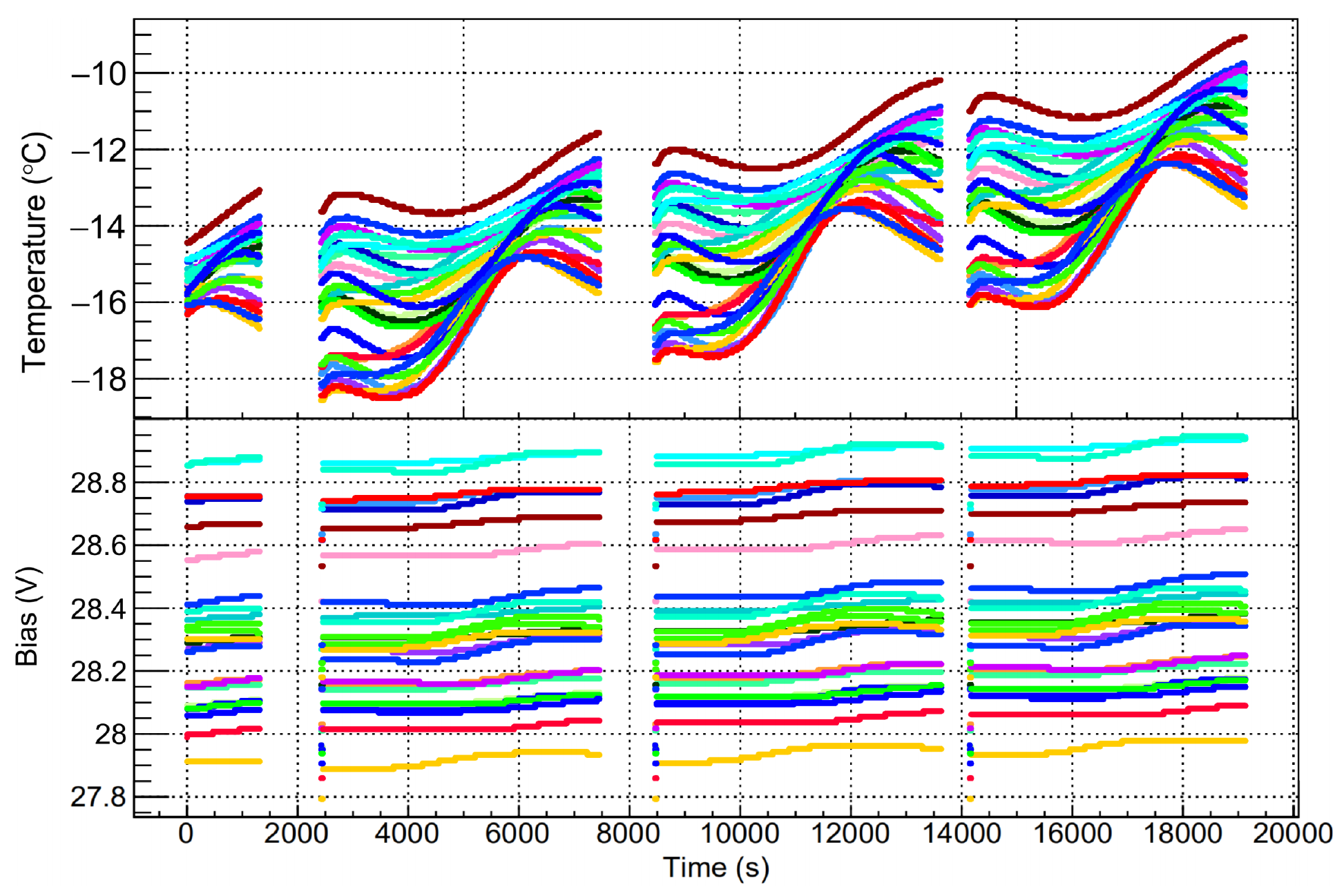}
  \hspace{0.1cm}
  \includegraphics[width=7.3 cm]{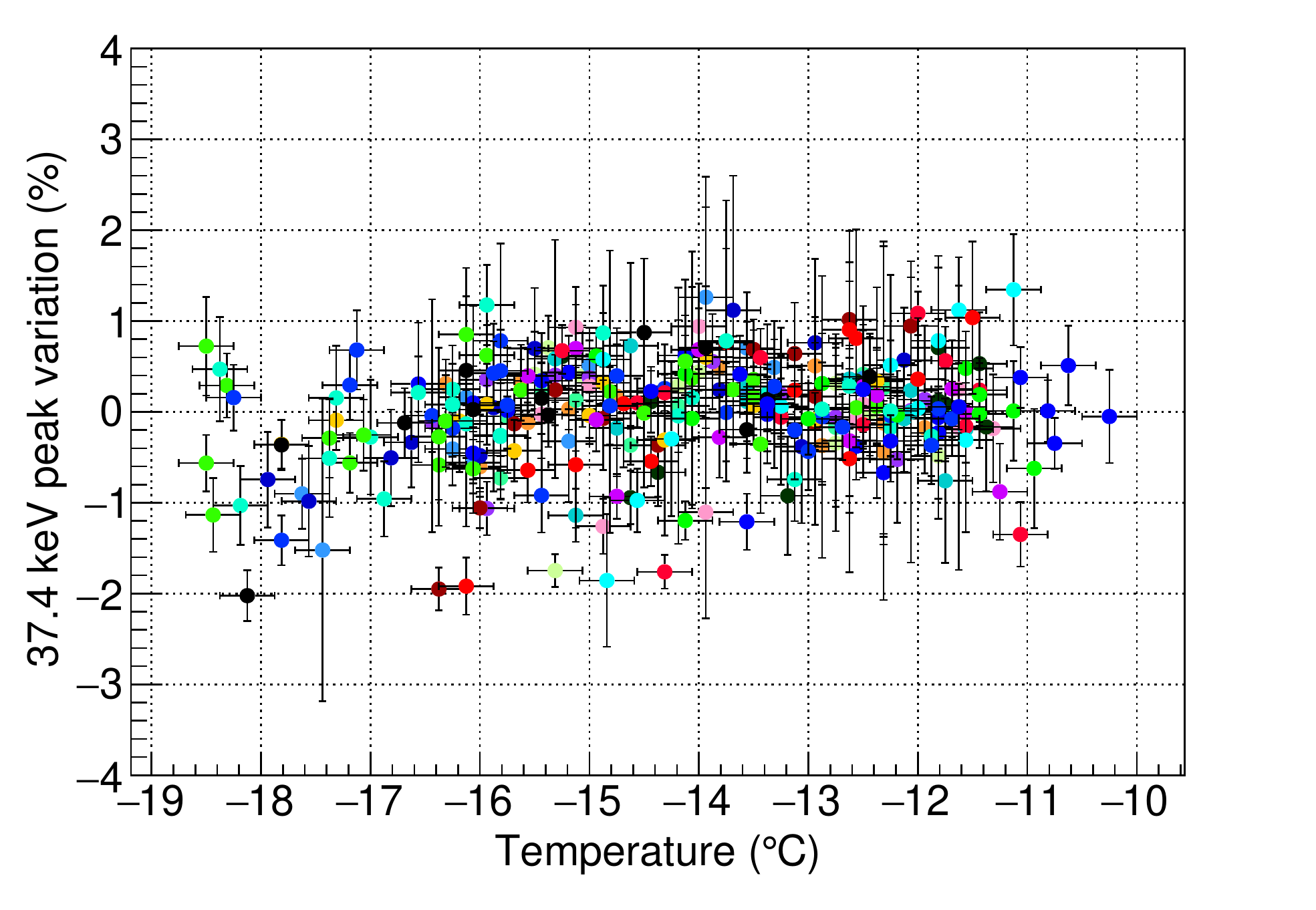}
  \caption{Left panel: In-flight temperature variation and real-time bias voltage adjustment. Right panel: Relative peak position variation from mean value VS temperature. The color represents different GRDs.}\label{Fig11}
  \vspace{0cm}
\end{figure}

 The designed working temperature of the GECAM GRDs is -20$\pm$3$^{\circ}$C with a temperature change rate that is within 3.3$^{\circ}$C/h. {\color{black}However, due to the satellite, GRD operates in an unexpected new mode. The} in-flight temperature variation is worse than expected. Therefore, the real-time SiPM bias adjustment is even more important to the GRD gain stabilization. {\color{black}Because the GRD gain is changed after launch, the initial bias voltage LUT (V10 version) is updated to V12 version with 2 iterations. Fig.\ref{Fig11} (left panel) shows the temperature and V12 bias voltage of 25 GRDs from UTC time 2021-01-15 02:00:00 to 07:19:00. It shows the bias voltage is automatically adjusted via some temperature variation.}
\par Some of the data is not available because the GECAM payload is shut down during the {\color{black}South Atlantic Anomaly (SAA)} area. Serval sets of spectrum data are accumulated for 4 minutes and the peak position variations of 1.47 MeV energy line are extracted. Fig.\ref{Fig11} (right panel) shows the relative peak position variations of the 25 GRDs. Most of the peak position variations are within 1.2\% whereas a few points are within 2\% in the temperature range of 18.5 $^{\circ}$C to -8 $^{\circ}$C. This result shows a good temperature stability under the real-time bias voltage adjustment.

\subsection{In-flight peak position consistency of characteristic lines}\label{section3.2}
\par
\begin{figure}[htbp]
  \centering
  \includegraphics[width=7 cm]{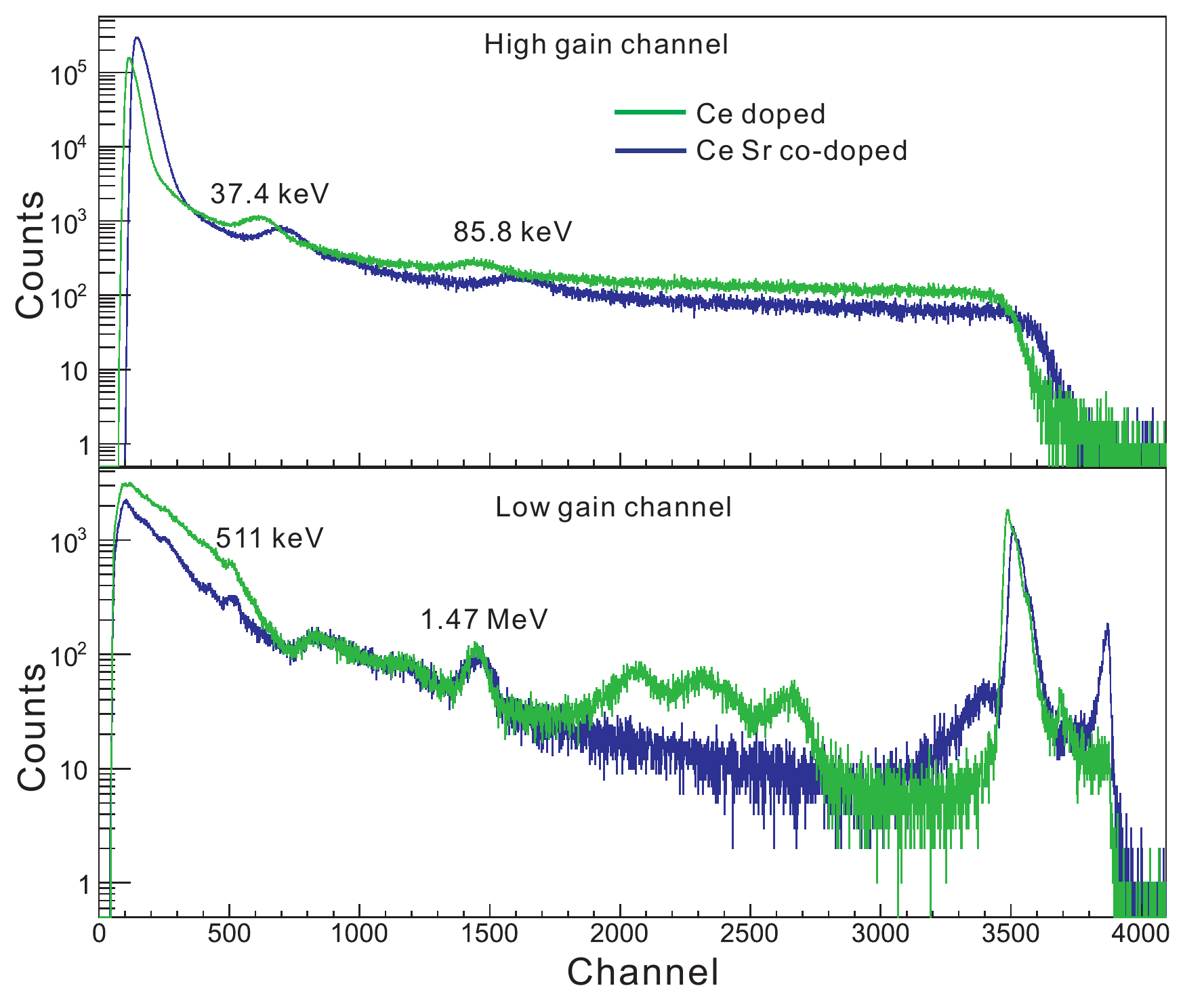}
  \hspace{0.1cm}
  \includegraphics[width=7.9 cm]{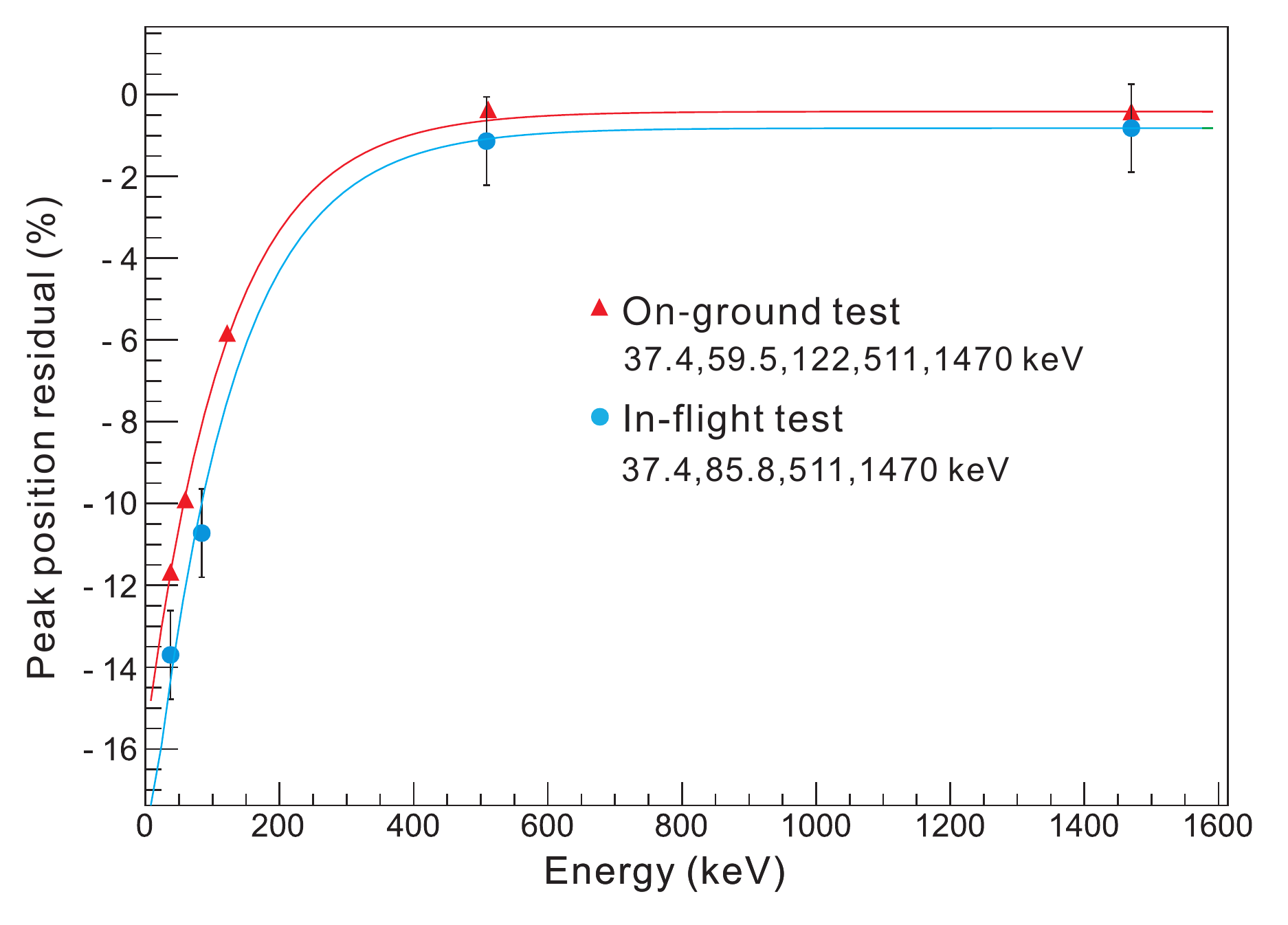}
  \caption{Left panel: In-flight background spectrum of single and co-doped GRDs. The cumulative time of the measurement is 1000 s. Right panel: Peak position residual between Ce doped GRD11 and co-doped GRD7.}\label{Fig12}
\end{figure}

 The in-flight background characteristic lines are important for the gain stabilization and adjustment of multiple GECAM gamma-ray detectors. The characteristic lines mainly consist of cosmic X-ray background, SAA proton activated LaBr$_{3}$ radiation, albedo gamma, cosmic proton and LaBr$_{3}$ inner radiation \cite{Inflight_BK}. The in-flight GRD gain consistency correction is performed by the approach mentioned in section \ref{section2.4}. The GRD non-uniformity under V11 and V12 bias voltage LUT are 4\% and 1.3\% respectively. Fig.\ref{Fig12} (left panel) shows the in-flight background of {\color{black}single doped (Ce only) and co-doped (Ce and Sr)} LaBr$_{3}$ under the V12 bias voltage LUT. The 37.4 keV and 1.47 MeV intrinsic gamma-ray lines of LaBr$_{3}$, galactic 511 keV gamma-ray line and activated 85.8 keV energy line are resolved from the background. The three peaks originating from the alpha decay of $^{227}$Ac and its daughters \cite{LaBrbackground} are around 2000$\backsim$3400 channels for a single doped GRD. These peaks of co-doped GRD are submerged by the saturated signal peaks (above 3400 channels) caused by the data acquisition system. After the consistency correction under V12 bias voltage LUT, peak position differences from the two types of GRDs are defined by the relative residuals (\textit{Res}) and fitted by an {\color{black}empirical} equation:
\begin{equation}
Res(\%)=100\cdot(\frac{P_{single}}{P_{co}}-1)=p_{0}+p_{1}\cdot e^{-(E-p_{2})/p_{3}}\label{fit}
\end{equation}
\par where \textit{P$_{single}$} and \textit{P$_{co}$}  is the peak position of single doped and co-doped GRDs, respectively. The red triangle markers are on-ground radioactive source test results and it shows obvious peak position residual change below 500 keV. By making use of the in-flight GRD characteristic lines, a similar peak position residual is shown via the blue square markers in Fig.\ref{Fig12} (right panel). {\color{black}The time of data sample is 1000 s.}
\par The in-flight data has larger error bars. This is caused by the gain drift in the temperature variation. In the fit process, the parameter \textit{p$_{0}$} is the peak position residual at 1.47 MeV. Parameter \textit{p$_{2}$} and \textit{p$_{3}$} are set to be 130 and 120, respectively. The fit parameters of on-ground test results are \textit{p$_{0}$}=-0.413, \textit{p$_{1}$}=5.214 and $\chi$$^{2}$/ndf is 2.445/4. The fit parameters of in-flight results are \textit{p$_{0}$}=-0.821, \textit{p$_{1}$}=6.238 and $\chi$$^{2}$/ndf is 1.046/3. Given a highly aligned peak position with non-uniformity around 1\%, the energy response of 25 GRDs can be categorized into two types. {\color{black}The peak position residual analysis shows that in the in-flight calibration process\cite{GECAM-In-flight-calibration}, different doped GRDs can use the same fitting parameters for energy to channel (E-C) relation of energy range above 511 keV. In the energy range lower than 511 keV, the in-flight calibration process need to take Eq.(\ref{fit}) into consideration for E-C relation correction between different doped GRDs. The further research on the peak position residuals will carry on in a dedicated paper\cite{GCAM-GRD-NPR}.}

\section{Conclusions and future work}
\par In this work, a novel approach for stabilizing and normalizing the gain of multiple GECAM GRDs has been proposed. The GRD and data acquisition system are enclosed in an active temperature feedback loop that automatically modifies the bias voltage to ensure the gain remains stable within 2\%. Moreover, although {\color{black}each} of the GRD has unique light yield, the gain consistency correction successfully reduced the non-uniformity from 17\% to 0.6\% via a specially designed bias voltage LUT. All the key parameters of the consistency correction approach are determined via on-ground tests and effectiveness of the proposed approach have been demonstrated via experimental validation. One of the main advantages of this approach is that,during the in-flight working period, the bias voltage adjustment strategy can be changed with the state of GECAM . Although uploading is available, the changes need to be very conservative to {\color{black}achieve} a relative better GRD performance. The iterative bias voltage adjustment principle delineated in this paper is highly effective and important.
\par The in-flight results of the multiple GRD adjustments provide important data that can be used for future development. The GECAM team is now developing a new gamma-ray detection payload called High Energy Burst Searcher (HEBS) that has a similar design to the GECAM. This new project is planned to be launched in early 2022. The HEBS GRD consists of NaI (TI) and LaBr$_{3}$ crystals. Because of the obvious temperature dependence of NaI crystal, follow-up research on temperature compensation approach for NaI typed GRD is underway. The gain stabilization and consistency approach of multiple SiPM based gamma-ray detectors will help to lay the foundation for future research in this field.

\section*{Acknowledgements}
\par We would like to express our appreciation to the staff of the National Institute of Metrology and Shandong Institute of Aerospace Electronic Technology who offer great help during the development phase. This research is supported by the Strategic Priority Research Program of Chinese Academy of Sciences, Grant No.XDA15360102, National Natural Science Foundation of China (12173038) and National Natural Science Foundation of China, Grant No.11775251. The authors would also like to thank the anonymous reviewers for their detailed and constructive comments in the evaluation of this paper.

\bibliographystyle{elsarticle-num}


\end{document}